\documentstyle[11pt,newpasp,twoside,epsf]{article}
\def\edcomment#1{\iffalse\marginpar{\raggedright\sl#1\/}\else\relax\fi}
\reversemarginpar

\include{texdefs}

\begin{document}
\title{Modeling Observational Signatures of Disk-Driven Outflows}
 \author{J.E. Everett, A. K\"{o}nigl, and J.F. Kartje}
\affil{Department of Astronomy and Astrophysics, University of Chicago,
5640 S. Ellis Avenue, Chicago, IL  60637}

\begin{abstract}
We present a self-consistent, semi-analytical dynamical model of disk
driven outflows in AGNs that are accelerated by a combination of
magnetic stresses and radiation pressure. This model will make it
possible to examine scenarios in which the wind is homogeneous as well
as cases where it consists of dense clouds embedded in a more tenuous,
magnetized medium. The various ingredients of this model will be
tested through quantitative predictions from both multiwavelength
spectral observations and reverberation mapping of AGNs.
\end{abstract}

\section{Introduction}

	Many researchers have theorized about the structure and
dynamics of the gas near the cores of AGNs, specifically in the Broad
Emission Line Region (BELR), Broad Absorption Line Region (BALR), and
the Warm Absorber (WA).  In addition, there are open questions as to
how all of these different regions might be related, if at all.

	In attacking these questions, different investigators have
invoked various processes and physical effects to explain observations
of AGNs.  Researchers have hypothesized that the gas may exist either
as discrete clouds or as continuous winds.  The idea that the gas is
partitioned into discrete clouds is the more ``traditional'' approach
to BELR modeling (Bottorff et al. 1997).  However, recent spectral
analysis by Arav et al. (1997, 1998) seem to show that a continuous
wind might be better suited to explain high resolution spectra of
broad lines.

	Second, to explain the acceleration of the gas to the
velocities observed in spectral lines, various researchers have
suggested either radiative acceleration or magneto-centrifugal
acceleration.  Radiative acceleration from a central continuum source
was proposed by Arav, Li, \& Begelman (1994) and Arav (1996) to
explain the acceleration of gas in the BALR.  Later, Proga (2000)
examined radiative acceleration due to continuum emitted by a
circumnuclear accretion disk as well as the central source.
Other researchers have invoked magneto-centrifugal acceleration
(K\"{o}nigl \& Kartje, 1994), which appears to explain the central
dusty torus through the natural vertical density stratification in the
magneto-centrifugal wind.

	Of course, investigators don't usually include just one of the
above possibilities; many different combinations of mechanisms have
been attempted. Emmering, Blandford, \& Shlosman (1992) modeled the
cores of AGNs by using a magneto-centrifugally driven continuous
outflow to simulate discrete clouds.  This model was later improved
(Bottorff et al. 1997) to explain line variabilities.  

	Working with radiative acceleration only, Murray et al. (1995)
and later Chiang \& Murray (1996) modeled the gas as a continuous
wind, driven by continuum and line radiation forces from both the
central source and the disk.  They used the inner edge of the wind as
a shield and therefore as a possible origin of the Warm Absorber, and
also explained the Broad Emission Line (BEL) profiles with the wind
itself. 

	Finally, de Kool \& Begelman (1995) used a continuous wind with
magneto-centrifugal driving and continuum radiation pressure to
examine the effects of radiation pressure on the magnetic outflow
structure.  Also, K\"{o}nigl et al. (1995) and Kartje et al. (1997)
explained EUV absorption features in BL Lac objects using both
magnetic and radiative acceleration.  They accomplish this by first
lifting material vertically away from the disk using
magneto-centrifugal driving (modified by including an approximate
radiation pressure in the self-similar wind equations).  The material
travels upward along the magnetic streamlines until it intercepts the
central beaming cone of the BL Lac, where the gas is forced out
radially by both line and continuum radiation pressure.  The clouds
move radially outward, absorbing the central continuum at a range of
velocities as they accelerate, yielding broad absorption troughs in BL
Lac spectra.

\section{Open Questions}
	As one might imagine, there are still a few open questions to
consider about many of these models.  

	Specifically, concentrating on the Murray et al. model, we
need to understand how well the radiative acceleration actually works
for a continuous wind as postulated on the scales and parameters in
this model.  In their paper, they cite approximate calculations to
show that it may be feasible, but no explicit model for the shielding
column is presented, although it is theorized that the shield is a
``failed wind'' region driven by disk radiation pressure that falls
back to the disk.  The difficulty is that the radiative acceleration
being proposed has never been simulated for the continuous wind
hypothesized in these models.  We point out that even though it has
been hypothesized that radiative acceleration from the disk continuum
could push material up into a wind (Proga 2000), a magneto-centrifugal
wind can also accomplish that, and can already explain the observed
obscuration near the disk.  In addition, as pointed out by Kartje et
al. (1997), such winds could be important for lifting material into
the beaming cone of BL Lacs.  Also, in BALQSO objects,
magneto-centrifugal winds could explain the more equatorial flow
pattern if the magnetic streamlines are pushed by radiation pressure
from being nearly vertical to being more radial or equatorial.  The
winds could then be the source of both the BL Lac and BALQSO absorption
features.  These options are displayed in Figures 1 and 2, where we
present a possibly geometry for BL Lac and BALQSOs with a
magneto-centrifugal wind and clouds.

\begin{figure}
\plotfiddle{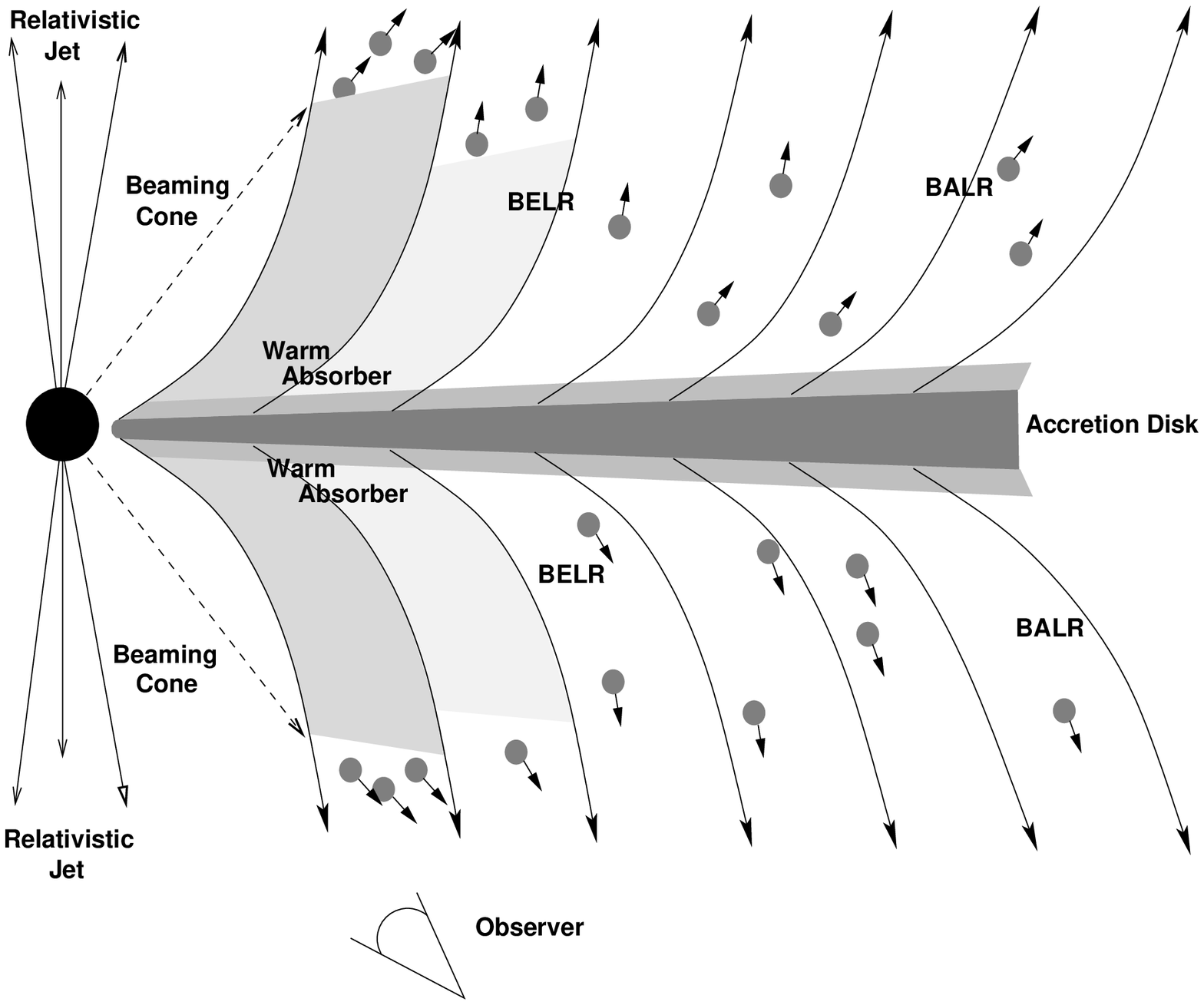}{3.0in}{0}{50}{50}{-160}{-100}
\caption{Schematic Geometry for a BL Lac with a magneto-centrifugal
wind and clouds.  The curved lines up from the disk represent magnetic
field lines.  The shielding column required for radiative
acceleration exists in the darker shaded regions above the accretion
disk as well as at the innermost portion of the wind.  In this class
of AGN, the magneto-centrifugal wind lifts the clouds out of the plane
of the accretion disk, where they intercept the central continuum and
are radiatively accelerated.}
\end{figure} 

\begin{figure}
\plotfiddle{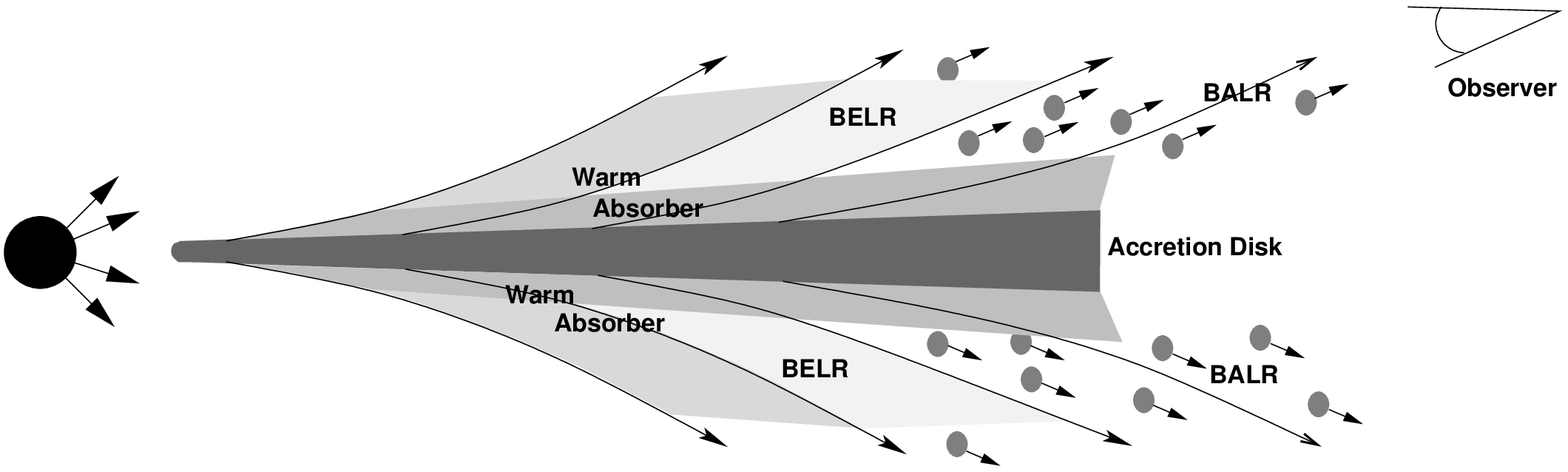}{1.9in}{0}{55}{55}{-165}{-150}
\caption{Possible Geometry for a BALQSO with a magneto-centrifugal
wind and clouds.  As in the BL Lac figure, the shielding column
required for radiative acceleration exists in the darker shaded
regions above the accretion disk as well as at the innermost portion
of the wind.  In BALQSOs, the magnetic field lines are pushed away
radially by the radiation pressure from the central source on the
clouds in the wind, resulting in a much more equatorial wind.}
\end{figure} 

	There are open questions left by the K\"{o}nigl \& Kartje work
as well.  First, how well can a ``clumpy'' medium explain both BL Lacs
and BALQSOs, as postulated above?  Second, are the BELR, BALR, and WA
regions present in their wind, or is it not possible to explain all of
these observed features with one model of this type?  The
magneto-centrifugal wind model accounts for the density stratification
near the disk, the uplifting of gas into the BL Lac beaming cone, and
would naturally produce the gas shield required in the Murray et
al. (1995) theory for radiative acceleration to operate.  Can this
wind then also produce a BELR within its inner region, perhaps?  Does
the outer region of the wind contain gas like that observed in the
BALR? 

	Third, and more generally, while the K\"{o}nigl \& Kartje work
postulates a ``clumpy'' medium, can a continuous medium explain the
observations without magnetic fields?  Is it possible that radiation
pressure from the disk could launch a continuous wind?  In this case,
radiation pressure would substitute for magnetic field acceleration
near the disk, and whereas magnetic fields are usually invoked to
confine clouds, with a continuous wind such confinement would not be
needed.  Also, while variability observations would seem to fit the
picture of a clumpy medium, Arav et al. (1999) suggest that even a
continuous medium could produce variability due to instabilities in
line-driven continuous winds.  Therefore, we ask, how can we test
observationally for a ``clumpy'' medium, and indeed, is a clumpy
medium required?  Can we constrain the gas to be clumpy or continuous
based on observations?
	
	We need a model to compare to the observations, both for the
effects of a clumpy or continuous medium and to compare radiative
acceleration vs. magnetic effects.  We are developing such a model,
and will use that model to generate predictions that will constrain the
physics at the cores of AGNs.

\section{Proposed Hybrid Model}
	In pursuit of answers to the above questions, we are
developing a quantitative, semi-analytical method that includes
many of the above effects to enable comparisons with observations for
multiple types of AGNs within a self-consistent model.  We include:

\begin{itemize}
\item a self-similar model of the magnetohydrodynamic wind,
\item radiation pressure from the central source,
\item the two-phase nature of the gas (clouds within a continuous gas
medium), including the coupling of the two phases, and
\item the effects of disk continuum (intrinsic and reprocessed) as a
source of radiative acceleration.
\end{itemize}

	We've chosen to start with these main ingredients because it
seems that we need magnetohydrodynamic, centrifugally driven winds and
radiation pressure to explain BL Lacs and perhaps BALQSOs, and
continuous winds have had other successes explaining broad emission
lines and reverberations results.  Therefore, we will combine them to
see what they produce when united, and what constraints their ``dual''
presence requires.

	In our progress so far, we have worked on implementing the
first two elements together.  We have developed a self-similar model
of the magnetohydrodynamic wind like that of K\"{o}nigl \& Kartje
(1994), except for an improved treatment of the radiative term,
$\Gamma(\theta)$ ($\Gamma(\theta) = a_{R}/g$, the ratio of the radial
radiative acceleration to the gravitational acceleration) in the
original equations.  This first program determines the shape of the
field lines from solving the cross-field (Grad-Shafranov) equation in
a self-similar formulation.  Next, we use the wind structure output
from that self-similar program as input to the photoionization code
Cloudy to determine the ionization state of the gas given the
self-similar scaling of the density.  The ionization output from
Cloudy is then used to more accurately compute the line and continuum
force multipliers that determine the radiative acceleration $a_{R}$:

\begin{equation}
a_{R} = \frac{n_{e} \sigma_{T} F}{\rho c} [M_{line}(\Xi,t) + M_{cont}(\Xi)],
\end{equation}
where $n_{e}$ is the electron density, $\sigma_{T}$ is the Thomson
scattering cross section, $F$ is the total incident radiation flux,
and $\rho$ is the density of the medium undergoing radiative
acceleration.  $M_{line}$ and $M_{cont}$ are the line and continuum
force multipliers, representing the ratio of line and continuum
acceleration to that of Thomson scattering alone.  The line force
multiplier is a function of $t$, the ``equivalent electron optical
depth scale'' (Arav, Li, \& Begelman 1994), related to the thermal
speed in the gas, $v_{th}$, $\epsilon$ and the velocity gradient
through $t = \sigma_{T} \epsilon n_{e} v_{th}/(dv/dr)$, where
$\epsilon$ is a factor to reduce the optical depth due to the small
filling fraction of the absorbing matter.  Both force multipliers are
functions of the ionization state of the gas, $\Xi$:

\begin{equation}
\Xi = \frac{F_{ion}}{n_{H} k T c},
\end{equation}
where $F_{ion}$ is the hydrogen ionizing flux in the incident
continuum and $n_{H}$ is the density of hydrogen in the gas.  

	Returning to our numerical model, Cloudy determines the
ionization state of the gas necessary to calculate these expressions
accurately.  We then integrate Euler's equation, calculating the
poloidal motion along the magnetic field lines fixed by the
self-similar code, but using the improved estimates of $a_{R}$ from
the Cloudy simulations.  This allows a new determination of the
velocity, density, and radiation pressure along the flow line which is
then input back into our first self-similar program.  We repeat the
process, iterating towards convergence of the effects of both magnetic
and radiation forces.  In both of these steps, we take the clouds to
be confined by the wind's (mostly magnetic) pressure, as suggested by
earlier authors.  Our work will advance the inter-relation between
clouds and the magnetic fields as we will consider the force of the
clouds (when radiatively accelerated) on the magnetic field structure,
as well.

	When the first two elements of the model have been completed
and tested, we will proceed to the third element of our model (the
two-phase nature of the gas), which has not been considered by previous
researchers.  We will include the effects of ram pressure of the
continuous wind on the clouds, and examine what effect that pressure
has observationally and what impact those forces have on launching of
clouds as well as their later dynamics.  The fourth part of our model
(including radiation pressure from the disk continuum) has been
considered by Proga (2000).  We will compare this effect to the already
simulated magnetic effects to assess the relative importance of the
different ingredients and make observable predictions for each.

	Overall, we wish to build a model capable of producing
predictions of how all of the above components relate.  Matching the
model with observations will allow us to see which of the different
components work in the quantitative models: are any of the elements
absolutely required or are there perhaps included effects that don't
fit with the observations?

	This model, however, should not be taken as a unification
scheme or as a single model that aims to explain all AGN phenomena.
Our present goal is instead to check the different physical effects
that have been invoked by different researchers and compare them with
AGN observations to constrain the key processes active in different
kinds of AGNs.

\section{Observational Possibilities}

	There are many questions which the above theoretical/numerical
framework would address.  Below, we consider a few of the important
ones we will concentrate on.

	First, this framework will show how a two-phase gaseous medium
would affect the MHD wind flow, and should allow us to explore how
those effects would be visible in observations of AGNs, particularly
in transfer functions from reverberation mapping.  If a two-phase
medium exists in the cores of AGNs, we might expect to see gas with
different ionization parameters in similar sections of the wind, hence
sharing the same kinematic signature.  This model will predict the
effect of those different ionization states on the transfer function
used in reverberation mapping.

	Also, this model should be able to show what differentiates
the BELR, BALR, and WA in the context of all of the included physical
effects.  If we analyze incoming observations against this more
inclusive model, perhaps we can start to relate the various
components: how do their densities, velocities, and origins compare,
for instance?

	We also ask how dynamical effects within the wind will affect
AGN spectra.  Perhaps we can explain some of the observed variability
by outbursts in this kind of radiatively driven, magneto-centrifugal
wind?

	Finally, we will predict spectral lines that might be
observable by {\it Chandra} or {\it XMM}.  The model will also
calculate continuum and line polarization from each model to test for
the presence of different dynamics in different AGN.  Kartje (1995)
has already studied continuum polarization effects, and this model
will allow us to advance that work and examine line polarization
effects as well.

\acknowledgements

We acknowledge the support of NASA grant NAG5-9063.  Also, we thank
Rita Sambruna for her guidance in BL Lac modeling during the summer of
1999.

\end{document}